\documentstyle[12pt]{article}
\newcommand{\db}[2]{\{#1#2\}^*}
\newcommand{\dbp}[2]{\{#1#2\}'}
\newcommand{\ds}{\displaystyle}
\newcommand{\be}{\begin{equation}}
\newcommand{\ee}{\end{equation}}
\newcommand{\rf}[1]{(\ref{#1})}
\newcommand{\f}[1]{\varphi_{#1}}
\newcommand{\pcd}[1]{\mathstrut\partial #1}
\renewcommand{\theequation}{\thesection.\arabic{equation}}
\makeatletter
\@addtoreset{equation}{section}
\makeatother
\title{Relativistic quantum mechanics in the einbein 
field formalism}

\vspace{2cm}

\author{Yu.S.Kalashnikova\thanks{e-mail:yulia@vxitep.itep.ru},
   A.V.Nefediev\thanks{e-mail:nefediev@vxitep.itep.ru}
\\ Institute of
Theoretical and Experimental Physics\\ 117259, 
Moscow, Russia}
\date{}

\begin{document}
\maketitle

\begin{abstract}

The system of two relativistic particles with einbein fields is
quantized as a constrained system.
A method of the introduction of the Newton--Wigner collective 
coordinate
is discussed in presence of different gauge fixing conditions.
Some arguments are involved in the favour of Lorents covariant 
gauge fixing
conditions.
\end{abstract}

\section{Introduction}

The motion of the spinless pointlike particle 
is described by 
the action
\be
S=\int_{\tau_i}^{\tau_f}Ld\tau,\hspace*{1cm}
L=-m\sqrt{\dot{x}^2},
\label{11}
\ee
where $\dot{x}_{\mu}=\frac{\partial x_{\mu}}{\partial\tau}$, 
and $\tau$ is
the parameter specifying the position of the particle along 
its world line.
The action \rf{11} is invariant under reparametrization 
transformation
$\tau\to f(\tau)$, which makes the theory \rf{11} be a 
constrained theory
in the Dirac sense \cite{dirac}, or a gauge theory. It can be 
more convenient
to rewrite the original Lagrange function \rf{11} in a 
different way introducing the
so-called einbein field $e$ \cite{brink}:\footnote{In 
what 
follows we
work with the field $\mu$, refering to it 
as to einbein field; 
one should
have in mind that the correct definition of einbein is 
$e=\frac{1}
{\mu}$.}
\be
L=-\frac{\dot{x}^2}{2e}-\frac{e m^2}{2}=-\frac{\mu\dot{x}^2}{2}-
\frac{m^2}{2\mu},\hspace{0.5cm}\mu=\frac{1}{e}.
\label{12}
\ee

The reparametrization transformations of \rf{12} have the form
\be
\begin{array}{ll}
x_{\mu}(\tau)\to x_{\mu}(f(\tau))&f(\tau_i)=\tau_i,\;
f(\tau_f)=\tau_f\\
{}&{}\\
\mu(\tau)\to\mu(f(\tau))\left/\dot{f}(\tau)\right.&
\end{array}
\label{13}
\ee

There is a lot of reasons to deal with the Lagrange function 
\rf{12} rather than
\rf{11}, starting with the formal ones: the action in the 
form \rf{12} appears
naturally in the Feynman-Schwinger representation for the 
propagator of the
relativistic particle \cite{fradkin}, and can be most 
straightforwardly
generalized for the case of spinning particle \cite{brink, 
fradkin}.
Besides, the Lagrange function \rf{12} is quadratic in 
velocities, so that the
velocity can be expressed explicitly in terms of canonical 
momentum
$p_{\mu}=\frac{\partial L}{\partial \dot{x}_{\mu}}$. Another 
advantage is
that while the Lagrange function \rf{11} is singular in the 
limit $m\to 0$, the
Lagrange function \rf{12} is not, and it is possible to 
formulate the
Hamiltonian approach for the massles particle.

The theories \rf{11} and \rf{12} are, of course, 
equivalent, 
and it can be
easily verified by writing the Eiler--Lagrange equation 
for the field $\mu$:
\be
\frac{\partial}{\partial\tau}\frac{\partial L}
{\partial\dot{\mu}}-
\frac{\partial L}{\partial\mu}=0,
\label{14}
\ee
which gives $\mu=m/\sqrt{\dot{x}^2}$. Substituting 
it into the Lagrange function
\rf{12}, one recovers the original form \rf{11}. 
There are 
two ways to arrange
for the Hamiltonian
setting of the theory \rf{12}. One may treat the einbein
field as a dynamical variable, and to define the 
corresponding 
canonical
momentum. Another way makes use of the fact that no time 
derivatives of
$\mu$ enter the Lagrange function; then the standard 
procedure of canonical
description is applied to the Lagrange function \rf{12} 
with the condition $\mu=
m/\sqrt{\dot{x}^2}$. If, however, one wishes to 
perform a transformation
of coordinates which involves the field $\mu$, 
one is left with the first
possibility only.

The simplest nontrivial example of such a situation 
is the problem of
centre--of--mass motion separation in the system of 
two non--interacting
relativistic particles, and the present paper is devoted 
to the canonical
description of this system with einbein fields involved. 
The paper is
organized as follows: in Section 2 we recollect the 
case of one particle
in the einbein field formalism. In Section 3 the Hamilton 
function and
constraints are written out for the case of two particles. 
Particular cases of
the gauge fixing are described in the 
Subsections 3.1 and 3.2,
and the concluding remarks are given in the final Section.

\section{One free particle}

The canonical momenta for the system \rf{12} are 
defined as
\be
p_{\mu}=\frac{\partial L}{\partial\dot{x}_{\mu}},
\hspace{1cm}\pi=
\frac{\partial L}{\partial\dot{\mu}}
\label{21}
\ee

It follows from equation \rf{21} that there is one 
primary constraint,
\be
\f{1}=\pi,
\label{22}
\ee
and, in accordance with the general procedure \cite{dirac}, 
the Hamilton
function
is
\be
H=H_0+\lambda\f{1},\hspace{0.5cm}H_0=-\frac{1}{2\mu}
(p^2-m^2).
\ee

The primary constraint \rf{22} generates the secondary one 
as a consequence of
the equation of motion,
\be
\f{2}=\{\f{1}H\}=\frac{1}{2\mu^2}(p^2-m^2),
\ee
where the Poisson brackets are defined to be
\be
\{AB\}=
\frac{\pcd{A}}{\pcd{p_{\mu}}}\frac{\pcd{B}}{\pcd{x_{\mu}}}-
\frac{\pcd{A}}{\pcd{x_{\mu}}}\frac{\pcd{B}}{\pcd{p_{\mu}}}+
\frac{\pcd{A}}{\pcd{\pi}}\frac{\pcd{B}}{\pcd{\mu}}-
\frac{\pcd{A}}{\pcd{\mu}}\frac{\pcd{B}}{\pcd{\pi}}.
\ee

t may be easily checked that no other constraints appear, 
and one has
\be
\{\f{1}\f{2}\}=0,
\ee
{\it i.e.} the constraints $\f{1}$ and $\f{2}$ are the 
first 
class ones. The
presence of the first class primary constraint $\f{1}$ 
means 
that there is
a gauge freedom associated with the reparametrization 
symmetry \rf{13}.
As there are no second class constraints, and the Poisson 
brackets for the
variables $x_{\nu}$ and $p_{\mu}$ have the canonical form,
$\{p_{\mu}x_{\nu}\}=g_{\mu\nu}$,
one may proceed in the way suggested by Dirac
\cite{dirac}: to postulate the operator valued commutator
\be
[\hat{p}_{\mu}\hat{x}_{\nu}]=-ig_{\mu\nu},\hspace*{0.5cm}
\hat{p}_{\mu}=
-i\frac{\pcd{}}{\pcd{x_{\mu}}}
\ee
and obtain, without imposing any gauge fixing conditions, 
the Klein--Gordon
equaion
\be
(p^2-m^2)\Psi=0,
\ee
which follows from the constraint $\f{2}$ treated as a 
weak one, or
as the equation for the wave function of the system.

Alternatively, the gauge can be fixed to 
establish the 
scale for $\tau$.
For example, the time--like gauge can be 
fixed by imposing 
the additional
primary constraint
\be
\f{3}=x_0+\tau,
\ee
identifying in such a way the evolution parameter $\tau$ 
with the
proper time of the particle.

It proves more convenient \cite{arm} to make the canonical 
transformation
of the variables, so that
\be
x'_0=x_0+\tau
\ee
and to impose the above constraint in the form
\be
\f{3}=x'_0
\ee
clearly getting rid of explicit dependence on time.

One can easily find the corresponding partition 
function to be
\be
F(x,p',\tau)=p'_{\mu}(x_{\mu}+\tau g_{\mu 0}),
\ee
and the modified Hamilton function becomes
\be
H'=H+\frac{\pcd{F}}{\pcd{\tau}}=H+p'_0.
\ee

With constraints $\f{3}$ and its secondary partner 
$\f{4}=\{\f{3}H'\}$
we arrive at the set of four constraints, all of the 
second class,
whereas the physical Hamilton function takes the 
familiar form
\be
H'=\sqrt{\vec{p}'^2+m^2}
\ee
on the constraints surface.

\section{Hamilton function and constraints 
in the system of two particles}

The Lagrange function for two non-interacting particles,
\be
L=
-\frac{\mu_1\dot{x}_1^2}{2}-\frac{m_1^2}{2\mu_1^2}-
-\frac{\mu_2\dot{x}_2^2}{2}-\frac{m_2^2}{2\mu_2^2},
\ee
is invariant under two independent reparametrization 
transformations of
the kind \rf{13}. To separate the cenre--of--mass 
motion we introduce the new
variables:
\be
\begin{array}{l}
x_{\mu}=x_{1\mu}-x_{2\mu},\hspace*{0.5cm}X_{\mu}=
\frac{\ds\mu_1}{\ds \mu_1+\mu_2}x_{1\mu}+
\frac{\ds\mu_2}{\ds\mu_1+
\mu_2}x_{2\mu}\\
{}\\
\zeta=\frac{\ds \mu_1}{\ds \mu_1+\mu_2},\hspace{0.5cm}
M=\mu_1+\mu_2.
\end{array}
\label{32}
\ee

In terms of these new variables the Lagrange function becomes
\be
L=-\frac{m_1^2}{2M\zeta}-\frac{m_2^2}{2M(1-\zeta)}-
\frac12M\left(\dot{X}-
\dot{\zeta}x\right)^2-\frac12M\zeta(1-\zeta)\dot{x}^2
\label{33}
\ee

The canonical momenta, defined as
\be
\begin{array}{ll}
P_{\mu}=\frac{\ds \pcd{L}}{\ds \pcd{\dot{X_{\mu}}}},&
p_{\mu}=\frac{\ds \pcd{L}}{\ds \pcd{\dot{x_{\mu}}}},\\
{}&{}\\
\Pi=\frac{\ds \pcd{L}}{\ds \pcd{\dot{M}}},&
\pi=\frac{\ds \pcd{L}}{\ds \pcd{\dot{\zeta}}},
\end{array}
\ee
give the Hamilton function
$$
H=H_0+\Lambda\Pi+\lambda\pi,
$$
\be
H_0=
\frac{\varepsilon_1^2}{2M\zeta}+
\frac{\varepsilon_2^2}{2M(1-\zeta)}-\frac{P^2}{2M},\\
\ee
$$
\varepsilon_1^2=m_1^2-p^2,\hspace*{0.5cm}
\varepsilon_2^2=m_2^2-p^2,
$$
and primary constraints
\be
\begin{array}{l}
\f{1}=\Pi,\\
\f{2}=\pi+\left(Px\right).
\end{array}
\ee

Note that, as the transformation \rf{32} mixes 
the space variables $x_1,\;x_2$
and einbein fields $\mu_1,\;\mu_2$, the Lagrange 
function \rf{33} contains
explicitly
the time derivative of the variable $\zeta$, and 
the constrain $\f{2}$ is
not as trivial as $\f{1}$. On the other hand, the 
transformation \rf{32}
involves the coordinates only; therefore the Poisson 
brackets are canonical:
$$
\{AB\}=
\frac{\pcd{A}}{\pcd{P_{\mu}}}\frac{\pcd{B}}{\pcd{X_{\mu}}}-
\frac{\pcd{A}}{\pcd{X_{\mu}}}\frac{\pcd{B}}{\pcd{P_{\mu}}}+
\frac{\pcd{A}}{\pcd{p_{\mu}}}\frac{\pcd{B}}{\pcd{x_{\mu}}}-
\frac{\pcd{A}}{\pcd{x_{\mu}}}\frac{\pcd{B}}{\pcd{p_{\mu}}}
\hspace*{2cm}
$$
\be
\hspace*{4cm}
+\frac{\pcd{A}}{\pcd{\Pi}}\frac{\pcd{B}}{\pcd{M}}-
\frac{\pcd{A}}{\pcd{M}}\frac{\pcd{B}}{\pcd{\Pi}}+
\frac{\pcd{A}}{\pcd{\pi}}\frac{\pcd{B}}{\pcd{\mu}}-
\frac{\pcd{A}}{\pcd{\mu}}\frac{\pcd{B}}{\pcd{\pi}}.
\label{37}
\ee

Using \rf{37}, the secondary constraints are calculated as
\be
\begin{array}{l}
\f{3}=\{\f{1}H\}=-\frac{\ds\varepsilon_1^2}{\ds 2M^2\zeta}
-\frac{\ds\varepsilon_2^2}{\ds 2M^2(1-\zeta)}+
\frac{\ds P^2}{\ds 2M^2}
\vspace*{0.2cm}\\
\f{4}=\{\f{2}H\}=-\frac{\ds\varepsilon_1^2}{\ds 2M\zeta^2}
+\frac{\ds\varepsilon_2^2}{\ds 2M(1-\zeta)^2}+\frac{\ds 
\left(pP\right)}
{\ds M\zeta(1-\zeta)}.
\end{array}
\ee

No other constraints arise, and since the constraints
$\f{1},\;\f{2},\;\f{3},\;\f{4}$
commute with each other they are the first class ones. 
As there are
two independent gauge invariances, we have two primary 
first class constraints
$\f{1}$ and $\f{2}$.

As the set of constraints contains physical and 
non--physical variables in
the mixed way, it is inconvenient to proceed {\it a' la}
Dirac without gauge fixing
at all and treating the constraints as weak conditions. 
In the next section
we
describe two of many possibilities of gauge fixing.

\subsection{Time--like gauge}
\renewcommand{\theequation}{\thesubsection.\arabic{equation}}
\setcounter{equation}{0}

The time--like gauge in both variables, $x_0$ and $X_0$, 
can be fixed in the
way similar to that used at the end of Section 2. 
The additional gauge fixing
constraints we are to impose have the form
\be
\f{5}=X_0+\tau,\hspace*{0.5cm}\f{6}=x_0
\label{311}
\ee
but the canonical transformation is needed only with 
respect to $\f{5}$
as $\f{6}$ does
not contain time from the very beginning.
\footnote{Hereafter primes at the
new variables and the Hamilton function are 
omitted for simplicity}

Primary constraints \rf{311} give rise to a couple 
of secondary ones,
\be
\f{7}=\{\f{5}H\}=\frac{P_0}{M}-1,\hspace*{0.5cm}\f{8}=
\{\f{6}H\}=
\frac{p_0}{M\zeta(1-\zeta)},
\ee
composing the set of eight second class constraints 
together with
$\f{1},\;\f{2},\;\f{3}$ and $\f{4}$.

The Hamilton function becomes
\be
H=M=P_0
\label{3155}
\ee
on the constraints surface.

The most economic way to exclude from the 
consideration the non--physical
variables, present in the theory because of the 
second class constraints,
is to evaluate these variables from the 
corresponding constraints,
to substitute them into the rest of constraints 
and into the Hamilton
function
and to change the Poisson brackets for the physical 
variables for the
so--called Dirac brackets defined as
\be
\{AB\}^*=\{AB\}-\sum_{a,b}\{A\f{a}\}C^{-1}_{ab}\{\f{b}B\},
\label{db}
\ee
where $C^{-1}$ is the inverse matrix with respect to 
$C$ constructed from the
Poisson brackets of the second class constraints 
being excluded:
\be
C_{ab}=\{\f{a}\f{b}\}.
\ee

The principle of correspondence will read now:
\be
\{AB\}^*=-i[AB],
\ee
where $[AB]$ indicates the quantum commutator.

Choosing as the physical variables the space 
components of the coordinates and
momenta we arrive at the following set of Dirac brackets:
\be
\begin{array}{l}
\db{P_i}{X_k}=\delta_{ik}\\
{}\\
\db{X_i}{p_k}=\frac{\ds p_iP_k}{\ds M^2}\\
{}\\
\db{X_i}{X_k}=\frac{\ds x_ip_k-x_kp_i}{\ds M^2}\\
{}\\
\db{X_i}{x_k}=\frac{\ds x_iP_k}{\ds M^2}+
\frac{\ds x_ip_k}{\ds M^2}
\left(
\frac{\ds 1-\zeta}{\ds\zeta}-\frac{\ds\zeta}
{\ds1-\zeta}\right)\\
{}\\
\db{p_i}{x_k}=\delta_{ik}-\frac{\ds P_iP_k}{\ds M^2}-
\frac{\ds P_ip_k}
{\ds M^2}
\left(\frac{\ds 1-\zeta}{\ds\zeta}-\frac{\ds\zeta}
{\ds 1-\zeta}\right)\\
{}\\
\db{P_i}{P_k}=\db{P_i}{p_k}=\db{P_i}{x_k}=
\db{p_i}{p_k}=\db{x_i}{x_k}=0
\end{array}
\label{316}
\ee

It is convenient to define the spin variable as
\be
S_{ik}=x_ip_k-x_kp_i
\ee

Then the following brackets can be obtained from \rf{316}:
\be
\begin{array}{l}
\db{X_i}{X_k}=\frac{\ds S_{ik}}{\ds M^2}\\
{}\\
\db{X_i}{S_{kl}}=\frac{\ds 1}{\ds M^2}\left(P_kS_{il}+
P_lS_{ki}\right)\\
{}\\
\db{S_{ik}}{S_{mn}}=
S_{in}\left(\delta_{km}-\frac{\ds P_kP_m}{\ds M^2}\right)+
S_{km}\left(\delta_{in}-\frac{\ds P_iP_n}{\ds M^2}\right)\\
{}\\
\hspace*{2.1cm}+S_{mi}\left(\delta_{kn}-\frac{\ds P_kP_n}
{\ds M^2}\right)+
S_{nk}\left(\delta_{im}-\frac{\ds P_iP_m}{\ds M^2}\right)
\end{array}
\label{317}
\ee

The brackets \rf{316} and \rf{317} are noncanonical, 
and the Hamilton
function \rf{3155} on the constrains surface in terms 
of $\vec{p}$ and
$\vec{P}$ is
\be
H=\sqrt{\vec{P}^2+{\cal E}_1^2+{\cal E}_2^2+2
\sqrt{{\cal E}_1^2{\cal E}_2^2+
(\vec{p}\vec{P})^2}},
\ee
where ${\cal E}_i=\sqrt{m_i^2+\vec{p}^2},\;(i=1,\;2)$.

The set of brackets \rf{317} is well--known 
\cite{pryce, nw, hr};
to bring it into
the canonical form the Newton--Wigner variables 
should be defined:
\be
\begin{array}{l}
Q_i=X_i-\frac{\ds S_{ik}P_k}{\ds E(M+E)},\hspace*{0.5cm}
E=\sqrt{M^2-\vec{P}^2}\\
{}\\
J_{ik}=r_ik_k-r_kk_i
\label{3175}
\end{array}
\ee
which commute as
\be
\begin{array}{l}
\db{Q_i}{Q_k}=0\\
{}\\
\db{Q_i}{J_{kl}}=0\\
{}\\
\db{J_{ik}}{J_{mn}}=
\delta_{km}J_{in}+\delta_{in}J_{km}+\delta_{kn}J_{mi}+
\delta_{im}J_{nk},
\end{array}
\label{318}
\ee
where the proper internal variables are
\be
\begin{array}{l}
\vec{k}=\vec{p}+\frac{\ds (\vec{p}\vec{P})\vec{P}}
{\ds E(M+E)}\\
{}\\
\vec{r}=\vec{x}+\frac{\ds (\vec{x}\vec{P})\vec{P}}
{\ds E(M+E)}+
\frac{\ds (\vec{x}\vec{P})\vec{k}}{\ds 
E\omega_i\omega_2}
\left(\omega_1-\omega_2-\frac{\ds (\vec{k}\vec{P})}
{\ds M}\right)\\
{}\\
M=\sqrt{\vec{P}^2+(\omega_1+\omega_2)^2},\hspace*{0.5cm}
\omega_1=\sqrt{m_1^2+\vec{k}^2},
\hspace*{0.3cm}
\omega_2=\sqrt{m_2^2+\vec{k}^2}
\end{array}
\label{320}
\ee
with the following Dirac brackets:
\be
\begin{array}{l}
\db{k_i}{r_k}=\delta_{ik}\\
{}\\
\db{k_i}{k_k}=\db{r_i}{r_k}=\db{k_i}{Q_k}=\db{r_i}{Q_k}=0
\end{array}
\label{321}
\ee

The physical Hamilton function expressed in terms of 
the variable $\vec{k}$
from \rf{320} takes the form:
\be
H=\sqrt{\vec{P}^2+\left(\sqrt{m_1^2+\vec{k}^2}+
\sqrt{m_2^2+\vec{k}^2}
\right)^2};
\label{322}
\ee
so with the brackets \rf{318} and \rf{321} the system 
is quantized setting
$\hat{P}_i=-i\frac{\pcd{}}{\pcd{Q_i}}$ and $\hat{k}_i
=-i\frac{\pcd{}}{\pcd{r_i}}$.

The internal momentum $\vec{k}$ can be expressed 
in terms of the single
particle variables $\vec{p}_1$ and $\vec{p}_2$ with 
the result familiar from
textbooks (see {\it e.g.} \cite{novog}):
\be
\vec{k}=\vec{p}_1-\frac{\vec{P}\omega_1(\vec{p}_1)}{E}
+\frac{(\vec{P}\vec{p}_1)\vec{P}}
{E(E+M)},\hspace*{0.5cm}\vec{P}=\vec{p}_1+\vec{p}_2
\label{323}
\ee

If one deals with 3--dimentional momenta $\vec{p}_1$ 
and $\vec{p}_2$, then
the substitution \rf{323} for the relative momentum 
$\vec{k}$ is the only
way to separate the centre--of--mass motion and to 
obtain the Hamilton
function in the form \rf{322}.

\subsection{$(Px)=0$ gauge}
\setcounter{equation}{0}

In this subsection we present another, in our opinion 
more elegant way of
gauge fixing. Instead of \rf{311} we 
impose the constraint
\be
\f{5}=(Px),
\label{411}
\ee
which is manifestly covariant. The secondary 
constraint generated by
primary constraint \rf{411} is
\be
\f{6}=\frac{(pP)}{M\zeta(1-\zeta)}
\label{412}
\ee
and is also a covariant one.

With this gauge fixing constraints $\f{5}$ 
and $\f{6}$ the subset
$\{\f{2},\;\f{4},\;\f{5},\;\f{6}\}$ is a 
second class one, while
$\f{1}$ and $\f{3}$ remain  commuting with anything else, 
and are still the
first class ones. It offers the possibility to 
define the preliminary Dirac
brackets as
\be
\{AB\}'=\{AB\}-\sum_{a,b}\{A\f{a}\}C^{-1}_{ab}\{\f{b}B\},
\label{dbp}
\ee
where
$$
C_{ab}=\{\f{a}\f{b}\},
$$
and indeces $a$ and $b$ take the value 2, 4, 5, 6
only. As now constraints  $\f{4}$ and $\f{6}$ are 
of the second class,
then at the constraints surface one has
\be
\zeta=\frac{\varepsilon_1}{\varepsilon_1+\varepsilon_2},
\label{zet}
\ee
the condition that holds strongly. Similarly, 
the condition $\pi=0$, which
follow from the second class constraints 
$\f{2}$ and $\f{5}$, is also a
strong one, and we eliminate these variables.

Defining the spin variable $S_{\mu\nu}=
x_{\mu}p_{\nu}-x_{\nu}p_{\mu}$ we
have the following set of preliminary Dirac 
brackets \rf{dbp}:
\be
\begin{array}{l}
\dbp{P_{\mu}}{X_{\nu}}=g_{\mu\nu}\\
{}\\
\dbp{X_{\mu}}{X_{\nu}}=\frac{\ds S_{\mu\nu}}{\ds P^2}\\
{}\\
\dbp{P_{\mu}}{P_{\nu}}=0\\
{}\\
\dbp{X_{\mu}}{S_{\alpha\beta}}=
\frac{\ds 1}{\ds P^2}\left(P_{\beta}S_{\alpha\mu}+
P_{\alpha}
S_{\mu\beta}\right)\\
{}\\
\dbp{S_{\mu\nu}}{S_{\alpha\beta}}=
S_{\mu\beta}\left(g_{\nu\alpha}-\frac
{\ds P_{\nu}P_{\alpha}}
{\ds P^2}\right)+
S_{\alpha\mu}\left(g_{\nu\beta}-\frac{\ds P_{\nu}
P_{\beta}}{\ds P^2}\right)\\
{}\\
\hspace*{2.1cm}+S_{\beta\nu}\left(g_{\mu\alpha}-
\frac{\ds P_{\mu}P_{\alpha}}{\ds P^2}\right)+
S_{\nu\alpha}\left(g_{\mu\beta}-\frac{\ds P_{\mu}
P_{\beta}}{\ds P^2}\right)
\end{array}
\label{413}
\ee
and
\be
\begin{array}{l}
\dbp{X_{\mu}}{p_{\nu}}=\frac{\ds p_{\mu}P_{\nu}}
{\ds P^2}\\
{}\\
\dbp{X_{\mu}}{x_{\nu}}=\frac{\ds x_{\mu}P_{\nu}}
{\ds P^2}\\
{}\\
\dbp{p_{\mu}}{x_{\nu}}=g_{\mu\nu}-\frac{\ds 
P_{\mu}P_{\nu}}{\ds P^2}\\
{}\\
\dbp{P_{\mu}}{p_{\nu}}=\dbp{P_{\mu}}{x_{\nu}}=
\dbp{p_{\mu}}{p_{\nu}}=
\dbp{x_{\mu}}{x_{\nu}}=0
\end{array}
\label{414}
\ee

The set of Dirac brackets \rf{413} is 
typical for the systems obeying the
condition $S_{\mu\nu}P_{\nu}=0$ \cite{pryce, hr}, 
the latter being
conveniently provided by constraint \rf{411} 
and its conjugated partner
\rf{412}. These brackets are noncanonical but 
covariant, and there exists a
rather smart way to make them canonical \cite{pronko}.

To this end we, following the method of \cite{pronko}, 
define the tetrade of
vectors associated with the four--vector $P_{\mu}$ as
\be
e_{0\mu}=\frac{P_{\mu}}{\sqrt{P^2}},\quad 
e_{i\mu}e_{j\mu}=-\delta_{ij},
\quad e_{0\mu}e_{j\mu}=0,
\ee
and introducing the Cristoffel symbols, which define 
the transport of the
tetrade,
\be
\Gamma_{ij\alpha}=e_{i\mu}\frac{\pcd}{\pcd 
P_{\alpha}}e_{j\mu},\quad
\Gamma_{0j\alpha}=e_{0\mu}\frac{\pcd}{\pcd 
P_{\alpha}}e_{j\mu}.
\label{4145}
\ee

Note that here the indeces $i,\;j$ are the 
tetrade, not the
three--vector ones. It can be shown that the quantity
\be
Z_{\mu}=X_{\mu}+\frac12S_{ij}\Gamma_{ij\mu},
\label{4285}
\ee
where $S_{ij}=e_{i\mu}e_{j\nu}S_{\mu\nu}$, is a 
properly commuting variable:
\be
\begin{array}{l}
\dbp{Z_{\mu}}{Z_{\nu}}=0\\
{}\\
\dbp{P_{\mu}}{Z_{\nu}}=g_{\mu\nu}
\label{429}
\end{array}
\ee

The proper internal variables are defined as the 
tetrade components of
$x_{\mu}$ and $p_{\mu}$:
\be
x_i=-e_{i\mu}x_{\mu},\quad p_i=-e_{i\mu}p_{\mu},
\ee
and commute in the canonical way:
\be
\begin{array}{l}
\dbp{x_i}{x_j}=\dbp{p_i}{p_j}=\dbp{Z_{\mu}}{x_i}=
\dbp{Z_{\mu}}{p_i}=
\dbp{Z_{\mu}}{S_{ik}}=0\\
\dbp{p_i}{x_j}=\delta_{ij}
\end{array}
\label{430}
\ee

The details of the tetrade formalism which allow 
to prove the
formulae \rf{429} and \rf{430} are given in the Appendix.

We are still left with two first class constraints, 
$\f{1}$ and $\f{3}$,
but now, having the Dirac brackets for the 
variables $P_{\mu}$, $Z_{\mu}$,
$x_i$ and $p_i$ in the canonical form, we can 
treat the trajectory
constraint which follows from the constraint 
$\f{3}$ with account of
\rf{zet}
\be
P^2-{\cal M}^2=0,\quad {\cal M}^2=
\left(\sqrt{m_1^2+p_i^2}+\sqrt{m_2^2+p_i^2}\right)^2,
\ee
as a weak condition, and apply the Dirac quantization 
procedure arriving at
the Klein--Gordon equation for the wave function:
\be
\left(\hat{P}^2-{\cal M}^2(\hat{p}_i)\right)\Psi=0,
\ee
where $\hat{P}_{\mu}=-i\frac{\pcd{}}{\pcd{Z_{\mu}}}$,
$\hat{p}_i=-i\frac{\pcd{}}{\pcd{x_i}}$.

As the internal gauge was fixed by means of covariant 
condition \rf{411},
no surprise that the generator of Lorentz transformations
$$
M_{\mu\nu}=X_{\mu}P_{\nu}-X_{\nu}P_{\mu}+S_{\mu\nu}=
\hspace*{6cm}
$$
\be
\hspace*{3cm}
Z_{\mu}P_{\nu}-Z_{\nu}P_{\mu}-\frac12S_{ij}
(\Gamma_{ij\mu}P_{\nu}-
\Gamma_{ij\nu}P_{\mu})+S_{\mu\nu}
\label{440}
\ee
commutes in the proper way
\be
\begin{array}{l}
\dbp{M_{\mu\nu}}{M_{\alpha\beta}}=g_{\nu\beta}
M_{\mu\alpha}-
g_{\nu\alpha}M_{\mu\beta}+g_{\mu\beta}M_{\alpha\nu}-
g_{\mu\alpha}M_{\beta\nu}\\
{}\\
\dbp{M_{\mu\nu}}{P_{\alpha}}=g_{\mu\alpha}
P_{\nu}-g_{\nu\alpha}P_{\mu}
\end{array}
\ee
so that the system of the preliminary Dirac 
brackets is Poincar{\'e}--covariant
at the classical as well as quantum level, 
while the coordinate $Z_{\mu}$ is
not a four--vector, as it follows from \rf{440}.

One can proceed further, and fix the gauge in 
the centre--of--mass variables
also. For example, the time--like gauge is 
fixed as described in Section~2.
Straightforward but rather irksome calculations 
lead to the following
expression for the Newton--Wigner coordinate 
expressed already not through the
tetrade components of spin but through convenient 
3--dimensional ones:
\be
Z_i=X_i+\frac{\ds S_{ik}P_k}{\ds H(E+H)},\quad 
H=\sqrt{E^2+\vec{P}^2}
\label{3150}
\ee
where $E$ is the energy of the system in the 
centre--of--mass frame.
The form \rf{3150} is the correct of the 
Newton--Wigner variable for a
system gauged by the condition $S_{\mu\nu}
P_{\nu}=0$ \cite{pryce, hr}.
Note that this Newton--Wigner coordinate certainly 
coincides with that
defined in
\rf{3175}, but to see it explicitly one is to 
express them both in terms of the
same variables, $\vec{x}_1,\;\vec{x}_2,\;\vec{p}_1$ and
$\vec{p}_2$ for example (see {\it e.g.} \cite{hr}).

\section{Discussion}
\renewcommand{\theequation}{\thesection.\arabic{equation}}

A pragmatical reader may ask the question: why to 
bother with powerful
machinery of Dirac brackets to obtain the result 
familiar from the first's
year textbooks? Apart from general consideration of 
inner beauty, there
might be more practical reasons in using the 
formalism of einbein fields.

Indeed, in relativistic quantum mechanics the 
problem of gauge fixing is
closely connected to the problem of eliminating of 
the relative time
variable whatever it means. We have demonstrated 
that the einbein formalism
provides the natural environment for solving 
this problem. It can be done
in more standard way by defining the 
three--dimensional Newton--Wigner
variables \rf{3175}, as well as in more 
sophisticated way of introducing
the covariant analogue of these variables \rf{4285}.
The latter procedure allows to retain the 
Poincar{\'e}--invariance of the
theory even after gauge fixing, and one cannot 
overestimate this feature.
The method suggested can be generalized for the 
case of interacting particles,
in particular when the interaction depends 
on velocities.

This work was supported by grant $N^{\underline{0}}$ 
96--02--19184
of the Russian Fundamental Research Foundation.
\vspace*{0.5cm}

{\parindent=0cm\Large\bf Appendix}
\vspace*{0.4cm}

{\parindent=0cm
For reference purpose we collect here the formulae 
from the paper
\cite{pronko} relevant for the definition of 
the variable $Z_{\mu}$ and
calculation of the Dirac brackets.} It follows 
from equation \rf{4145} that
the tetrade vectors $e_{i\mu}$ and $e_{0\mu}$ 
satisfy the relations 
$$
\begin{array}{l}
\frac{\ds\pcd{e_{i\mu}}}{\ds\pcd{P_{\alpha}}}=
\Gamma_{ij\alpha}e_{j\mu}-
\Gamma_{i0\alpha}e_{0\mu},\\
{}\\
\frac{\ds\pcd{e_{0\mu}}}{\ds\pcd{P_{\alpha}}}=
\Gamma_{0i\alpha}e_{i\mu}.
\end{array}
\eqno{(A.1)}
$$

Treating these relations as a system of 
equations for $e_{i\mu}$ and
$e_{0\mu}$, one obtains:
$$
\begin{array}{l}
\frac{\ds\pcd{\Gamma_{0i\alpha}}}{\ds
\pcd{P_{\beta}}}-
\frac{\ds\pcd{\Gamma_{0i\beta}}}{\ds
\pcd{P_{\alpha}}}+
\Gamma_{ij\beta}\Gamma_{0i\alpha}-\Gamma_{ij\alpha}
\Gamma_{0i\beta}=0\\
{}\\
\frac{\ds\pcd{\Gamma_{im\beta}}}{\ds\pcd{P_{\alpha}}}-
\frac{\ds\pcd{\Gamma_{im\alpha}}}{\ds\pcd{P_{\beta}}}-
\Gamma_{ij\alpha}\Gamma_{jm\beta}+\Gamma_{ij\beta}
\Gamma_{jm\alpha}+
\Gamma_{i0\alpha}\Gamma_{0m\beta}-\Gamma_{i0\beta}
\Gamma_{0m\alpha}=0
\end{array}
\eqno{(A.2)}
$$
as the conditions under which the system $(A.1)$ 
has nontrivial solutions.
The tetrade vectors depend only on $P_{\mu}$, 
and therefore
$$
\begin{array}{l}
\dbp{X_{\mu}}{e_{i\nu}}=-\frac{\ds\pcd{e_{i\nu}}}
{\ds\pcd{P_{\mu}}}\\
{}\\
\dbp{X_{\mu}}{e_{0\nu}}=-\frac{\ds\pcd{e_{0\nu}}}
{\ds\pcd{P_{\mu}}}\\
{}\\
\dbp{X_{\mu}}{\Gamma_{ij\nu}}=-\frac{\ds\pcd
{\Gamma_{ij\nu}}}{\ds\pcd{P_{\mu}}}
\end{array}
\eqno{(A.3)}
$$

With the help of commutators \rf{413} 
the relations $(A.1)-(A.3)$ give
formulae \rf{429} and \rf{430}.

The explicit choice of the tetrade is not unique 
as it is clearly seen from
its definition. In the given reference frame 
the tetrade may be chosen, for
example, as
$$
e_{0\mu}=\frac{P_{\mu}}{\sqrt{P^2}},
\quad e_{i0}=\frac{P_i}{\sqrt{P^2}},\quad
e_{ik}=\delta_{ik}+\frac{P_{i}P_{k}}{\sqrt{P^2}
(P_0+\sqrt{P^2})}
\eqno{(A.4)}
$$
\textheight 190cm


\begin{thebibliography}{99}
\bibitem{dirac}P.A.M.Dirac, 
"Letures on Quantum Mechanics", Belter
Graduate\\
School of Science, Yeshiva University, 
New York (1964)
\bibitem{brink}L.Brink, P.Di Vecchia, 
P.Howe, Nucl.Phys. {\bf B118} (1977) 76
\bibitem{fradkin}E.S.Fradkin and D.M.Gitman, 
Phys.Rev. {\bf D44} (1991) 3230
\bibitem{arm}G.V.Grygoryan and R.P.Grigoryan, 
Yad.Fiz. {\bf 53} (1991) 1737
(in Russian)
\bibitem{pryce}M.H.L.Pryce, Proc.Roy.Soc. London, 
Ser.{\bf A150} (1935) 166
\bibitem{nw}T.D.Newton and E.P.Wigner, Rev.Mod.Phys. 
{\bf 21} (1949) 400
\bibitem{hr}A.J.Hanson and T.Regge, Ann.Phys. 
{\bf 87} (1974) 498
\bibitem{novog}Yu.V.Novozhilov 
"Vvedenie v teoriyu elementarnyh chastits",
Moskva, "Nauka" (1972) (in Russian)\samepage
\bibitem{pronko}G.P.Pron'ko, A.V.Razumov, 
Theor.Math.Phys. {\bf 56} (1983) 192
(in Russian)
\end{thebibliography}
\end{document}